\documentclass[superscriptaddress,twocolumn,showpacs,amsmath,amssymb]{revtex4}

\usepackage{graphicx}

\renewcommand{\vec}[1]{\mbox{\boldmath $#1$}}

\begin{document}

\title{
Three-body model calculations for $^{16}$C nucleus} 

\author{K. Hagino}
\affiliation{ 
Department of Physics, Tohoku University, Sendai, 980-8578,  Japan} 

\author{H. Sagawa}
\affiliation{
Center for Mathematical Sciences,  University of Aizu, 
Aizu-Wakamatsu, Fukushima 965-8560,  Japan}


\begin{abstract}
We apply a three-body model consisting of two valence neutrons and the
core nucleus $^{14}$C in order to investigate the ground state properties 
and the
electronic quadrupole transition of the $^{16}$C nucleus. 
The discretized continuum 
spectrum within a large box is taken into account 
by using a single-particle basis obtained from a Woods-Saxon potential. 
The calculated  B(E2) value from the first 2$^+$ state to the ground state
shows good agreement with the observed data with  the core 
polarization charge which reproduces the
experimental B(E2) value for $^{15}$C.
We also show that the present calculation well accounts for 
the longitudinal momentum distribution of $^{15}$C fragment 
from the breakup of $^{16}$C nucleus. We point out that 
the dominant ($d_{5/2})^{2}$ configuration in the ground state of 
$^{16}$C plays a crucial role for 
these agreement. 

\end{abstract}


\maketitle

Nuclei far from the $\beta$ stability line often reveal unique
phenomena originating from a large asymmetry in the neutron and 
proton numbers. One of the typical examples is a neutron 
collective mode, 
which is characterized only by the neutron 
excitation, with negligible contribution from the proton excitation. 
A recent calculation based on the 
continuum quasi-particle random phase approximation (QRPA) 
has, in fact,   predicted the existence of such neutron mode 
in the low-lying quadrupole excitation in $^{24}$O \cite{M01}. 

Recently, the electric quadrupole (E2) transition from the first 
2$^+$ state at 1.766 MeV to the ground state in $^{16}$C has been 
measured at RIKEN \cite{IOA04}. 
The observed B(E2) value (0.26$\pm$0.05 Weisskopf units) 
has turned out to be 
surprisingly small, as compared to the known systematics in 
stable nuclei. 
On the other hand, a distorted wave Born approximation 
(DWBA) analysis for the $^{16}$C + $^{208}$Pb inelastic scattering 
indicates a large enhancement of the ratio of the neutron to 
proton transition amplitudes, $M_n/M_p$=7.6$\pm$1.7 \cite{EDK04}, 
that is considerably larger than the isoscalar value, $N/Z$=1.67. 
A similar value for $M_n/M_p$ was also obtained from the 
inelastic proton scattering from the $^{16}$C nucleus \cite{OIA06}. 
These experimental data suggest that the first 2$^+$ state in $^{16}$C 
is a good candidate for the neutron excitation mode. 

There have already been various  theoretical calculations for the structure of 
$^{16}$C nucleus \cite{KE05,SZZT04,TIOI04,SMA04,HY06,FMOSA06}. 
Except for a recent microscopic shell-model calculation
\cite{FMOSA06}, however, they all fail to reproduce the anomalously
hindered E2 transition. For instance, 
Suzuki and his collaborators have solved a $n+n+^{14}$C three-body
model and found that the E2 strength is
overestimated by a factor of about 2 if the same core polarization
charge is employed as that used to describe the $^{15}$C nucleus. 
A similar overestimation of B(E2) value was found also in the 
antisymmetrized molecular dynamics (AMD) calculation \cite{KE05} 
as well as in the deformed Skyrme Hartree-Fock calculation
\cite{SZZT04}. 

In this paper, we apply a three-body model with a finite-range 
$n$-$n$ interaction 
 \cite{SMA04,HY06} to describe the ground and excited states in 
the $^{16}$C nucleus. 
We employ the single-particle (s.p.) basis obtained from 
a $n$-$^{14}$C Woods-Saxon potential to diagonalize the three-body Hamiltonian. 
The continuum s.p. spectrum is discretized in a large box. 
Notice that the effect of continuum couplings 
can be properly accounted for with such s.p. basis \cite{cd}. 
A similar three-body model with a density-dependent 
contact interaction has successfully been applied to describe the
structure of Borromean nuclei \cite{BE91,EBH99,VMP96,HS05,HSCS06}. 
In Refs. \cite{SMA04,HY06}, 
Suzuki {\it et al.} adopted the correlated Gaussian basis to diagonalize
a similar three-body Hamiltonian for  $^{16}$C.  
However, it remains an open question whether the correlated Gaussian basis 
is efficient enough to take into account the continuum couplings. 
Therefore, our study can 
be considered as a complement to the previous
studies in Refs.\cite{SMA04,HY06}.

Assuming that the effect of core excitation on the low-lying spectrum of the
$^{16}$C nucleus is negligible \cite{SMA04,VM95}, we 
consider the following three-body
Hamiltonian: 
\begin{equation}
H=\hat{h}(1)+\hat{h}(2)+V_{nn}+\frac{\vec{p}_1\cdot\vec{p}_2}{A_cm},
\label{3bh}
\end{equation}
where $m$ and $A_c$ are the nucleon mass and the mass number of the
inert core nucleus, respectively. 
$\hat{h}$ is the s.p. Hamiltonian for a valence
neutron interacting with the core.
The diagonal component of the recoil kinetic energy of the core
nucleus is included in
$\hat{h}$, whereas the off-diagonal part is taken into account
in the last term in the Hamiltonian (\ref{3bh}).
We use a Woods-Saxon potential for the interaction in $\hat{h}$, 
\begin{equation}
V_{nC}(r)=\left(V_0+V_{ls}\,(\vec{l}\cdot\vec{s})
\frac{1}{r}\frac{d}{dr}\right)
\left[1+\exp\left(\frac{r-R}{a}\right)\right]^{-1},
\label{WS}
\end{equation}
where $R=r_0A_c^{1/3}$.
The parameter sets for the Woods-Saxon potential 
which we employ in this paper are listed in Table I. 
The sets A, B, and C were used in Ref. \cite{SMA04,HY06}, while the set
D was used in Ref. \cite{VM95} in order to discuss the role of
particle-vibration coupling in the $^{15}$C nucleus. 
These parameter sets yield almost the same value for the energy of 
the 2$s_{1/2}$ state, $\epsilon_{2s_{1/2}}\sim -1.21 $ MeV, and of the 
the 1$d_{5/2}$ state, $\epsilon_{1d_{5/2}}\sim -0.47 $ MeV. 

\begin{table}[hbt]
\caption{
Parameters for the Woods-Saxon neutron-core potential, $V_{nC}$, in
Eq. (\ref{WS}). }
\begin{center}
\begin{tabular}{c|cccc}
\hline
\hline
Set & $V_0$ (MeV) & $V_{ls}$ (MeV fm$^2$) & $r_0$ (fm) & $a$ (fm) \\
\hline
A & $-$50.31 & 16.64 & 1.25 & 0.65 \\
B & $-$50.31 ($l$=0) & 31.25 & 1.25 & 0.65 \\
  & $-$47.18 ($l\neq 0$) &  &  &  \\
C & $-$51.71 & 26.24 & 1.20 & 0.73 \\
D & $-$44.41 & 31.52 & 1.27 & 0.90 \\
\hline
\hline
\end{tabular}
\end{center}
\end{table}

In our previous works \cite{HS05,HSCS06}, we used the
density-dependent delta force \cite{BE91,EBH99} 
for the interaction between the valence neutrons, $V_{nn}$. 
However, we here use the same finite-range force as in Ref. \cite{SMA04} 
in order to compare our results with those of Refs. \cite{SMA04,HY06}. 
That is the singlet-even part of the  Minnesota potential
\cite{Minnesota}, 
\begin{equation}
V_{nn}(\vec{r}_1,\vec{r}_2)=v_0\,e^{-b_0(\vec{r}_1-\vec{r}_2)^2}
+v_1\,e^{-b_1(\vec{r}_1-\vec{r}_2)^2},
\label{Minnesota}
\end{equation}
with $v_0$=200 MeV, $b_0$=1.487 fm$^{-2}$, and $b_1$=0.465 fm$^{-2}$. 
Following Ref. \cite{SMA04}, we adjust the value of $v_1$ for each
parameter set of the Woods-Saxon potential so that the ground state 
energy of $^{16}$C, $E_{\rm gs}=-5.47$ MeV, is reproduced. 

\begin{table}[hbt]
\caption{Properties of the ground and the second 0$^+$ states obtained
with several parameter sets for the neutron-core potential, 
$V_{nC}$. $P_{ss}$ and $P_{dd}$ are the
probability for the $[(2s_{1/2})^2]$ and $[(1d_{5/2})^2]$ components
in the wave function, respectively. 
$P_{S=0}$ is the probability of the 
spin-singlet ($S=0$) component in the ground state. 
$E_{0^+_2}$ is the excitation 
energy for the
second 0$^+$ state in the unit of MeV, while $r(^{16}$C) is the 
root-mean-square radius of 
the $^{16}$C nucleus in the unit of fm. The experimental values  
are  $E_{0^+_2}({\rm exp})=$3.00 MeV and 
$r(^{16}$C; exp)=2.64 $\pm$ 0.05 fm \cite{Z02}, respectively. } 
\begin{center}
\begin{tabular}{c|cccc|ccc}
\hline
\hline
Set & $P_{ss}$(g.s.) & $P_{dd}$(g.s.) & $P_{S=0}$ & $r(^{16}$C) & $E_{0^+_2}$ 
&
$P_{ss}$(0$_2^+$) & $P_{dd}(0_2^+)$ \\ 
\hline
A & 0.184 & 0.699 & 0.784 & 2.56 & 2.32 & 0.755 & 0.201 \\
B & 0.177 & 0.711 & 0.746 & 2.56 & 2.35 & 0.775 & 0.187 \\
C & 0.183 & 0.696 & 0.768 & 2.57 & 2.39 & 0.763 & 0.196 \\
D & 0.206 & 0.633 & 0.808 & 2.64 & 2.48 & 0.733 & 0.221 \\
\hline
\hline
\end{tabular}
\end{center}
\end{table}

The  three-body Hamiltonian (\ref{3bh}) 
is diagonalized 
by expanding the two-particle wave function
$\Psi(\vec{r}_1,\vec{r}_2)$
with the eigenfunction
$\phi_{nljm}$ of the s.p. 
Hamiltonian $\hat{h}$, where $n$ is the radial quantum number. 
The continuum s.p. states are discretized with a box size of 
$R_{\rm box}=30$ fm. 
We include the s.p. angular momentum $l_1$ and $l_2$ up to 5, and 
truncate the model space of the two-particle states at
$\epsilon_1+\epsilon_2$ = 30 MeV, where $\epsilon$ is the  s.p. energy 
of the valence particle. 
We have checked that the results do not significantly change 
even if we truncate the model space at 60 or 80 MeV, as long as $v_1$ in
Eq. (\ref{Minnesota}) is adjusted for each model space. 
In the diagonalization, we explicitly exclude the 
$1s_{1/2}, 1p_{3/2}$, and  $1p_{1/2}$ states, which 
are occupied by the core nucleus. 
The results for the ground state and the second 0$^+$ 
state are summarized in Table II. 
The parameter set dependence is small, although the set D 
reproduces the excitation 
energy of the second 0$^+$ state, $E_{0_2^+}$, and the
root-mean-square (rms) radius of the 
$^{16}$C nucleus, $r(^{16}$C), slightly better than the other parameter sets. 
The latter quantity is calculated as \cite{BE91,EBH99,VMP96} 
\begin{equation}
\langle r^2\rangle_{A_c+2}
=\frac{A_c}{A_c+2}\langle r^2\rangle_{A_c} 
+\frac{1}{A_c+2}\,\left(
\frac{2A_c\,\langle \rho^2 \rangle}{A_c+2}
+\frac{\langle \lambda^2 \rangle}{2}\right),
\end{equation}
where $\vec{\lambda}=(\vec{r}_1+\vec{r}_2)/2$ and
$\vec{\rho}=\vec{r}_1-\vec{r}_2$. Following Refs. \cite{SMA04,HY06}, 
we take 2.35 fm for the rms radius of the core nucleus, 
$\sqrt{\langle r^2\rangle_{A_c}}$. 
We find that the rms radius of $^{16}$C is well reproduced in the 
present calculations. 

We notice that our results are considerably different from those of 
Refs. \cite{SMA04,HY06} concerning 
the probability for the $[(2s_{1/2})^2]$ and $[(1d_{5/2})^2]$ components
in the wave function, which are denoted by $P_{ss}$ and $P_{dd}$ in Table II,
respectively. Our results show that the ground state of $^{16}$C 
mainly consists of the $[(1d_{5/2})^2]$ configuration, while the
second 0$^+$ state is dominated by 
 the $[(2s_{1/2})^2]$ configuration. This is in
contrast to the results of Refs. \cite{SMA04,HY06}, which show the 
dominance of the $[(2s_{1/2})^2]$ component in the ground state. 
As a consequence, we also obtain a smaller value of the spin-singlet 
probability, $P_{S=0}$, than in Ref. \cite{HY06}. 
Notice that the d-wave dominance was suggested from the analyses of 
longitudinal momentum distribution for the one-neutron knockout
reaction of $^{16}$C \cite{MAB01,Y03}. 
We will discuss the longitudinal momentum distribution later in this
paper.

It is worthwhile to consider a simple two-level pairing model 
consisting of the 2$s_{1/2}$ and 1$d_{5/2}$ s.p. levels in order to
illustrate how the $[(1d_{5/2})^2]$ configuration becomes 
dominant in the ground state of $^{16}$C. 
If there were no
interaction between the valence neutrons, the ground state wave
function would be the pure $[(2s_{1/2})^2]$ state, since the s.p. energy for
the 2$s_{1/2}$ state is lower than that for the 1$d_{5/2}$ state 
($\epsilon_{2s_{1/2}}=-1.21 $ MeV and $\epsilon_{1d_{5/2}}=
-0.47 $ MeV) . If one assumes a delta interaction,
$V_{nn}=-g\,\delta(\vec{r}_1-\vec{r}_2)$ between the valence neutrons, 
the diagonal matrix element of the Hamiltonian reads \cite{BB05} 
\begin{equation}
H_{ii}=2\epsilon_i-g\,\frac{2j+1}{8\pi}\,I_{ii},
\end{equation}
where $I_{ii}$ is the radial integral for the configuration
$(i)^2$. Therefore, the pairing interaction influences the 
$[(1d_{5/2})^2]$ configuration more strongly than the 
$[(2s_{1/2})^2]$ configuration by a factor of 3 when the radial 
integral $I_{ii}$ is similar to each other. 
If we choose the strength $g$ so that the ground state energy is
reproduced within the two-level model, we find $g$=1005
MeV$\cdot$fm$^{-3}$ for the parameter set D. 
This leads to the diagonal matrix element of 
$H_{ii}=-3.70$ MeV for $i=2s_{1/2}$ and 
$H_{ii}=-4.85$ MeV for $i=1d_{5/2}$, lowering the 
$[(1d_{5/2})^2]$ configuration in energy. Taking into account the 
off-diagonal matrix element and diagonalizing the 2$\times$2 matrix, 
we find $P_{ss}$=0.26 and $P_{dd}$=0.74
for the ground state, which are very close to the  results shown in 
Table II. 

The upper panels of Figs. 1 and 2 show the two-particle densities
$\rho_2(r_1,r_2,\theta)$ 
for the ground state
and the second 0$^+$ state, respectively. 
These are obtained with the Minnesota potential
and the parameter set D for the s.p. potential.  
In order to facilitate the presentation, we set $r_1=r_2=r$ and 
multiply the weight factor of $8\pi^2 r^4 \sin\theta$
\cite{HS05}. Despite that $P_{ss}$ and $P_{dd}$ are considerably
different, we obtain similar density distributions to those 
in Ref. \cite{HY06}. In particular, we observe similar ``di-neutron'' and
``cigar-like'' configurations in the ground state, as well as 
``boomerang'' configuration \cite{HY06} in the second 0$^+$ state 
as in Ref. \cite{HY06}. The lower panels of Figs. 1 and 2 show  the
angular densities $\rho(\theta)$ obtained by integrating the radial
coordinates in the two-particle density \cite{HS05}. It is multiplied
by a weight factor of 2$\pi\sin\theta$. As a comparison, we also
show the angular densities for the pure 
$[(2s_{1/2})^2]$ and 
$[(1d_{5/2})^2]$ configurations by the dotted and dashed lines,
respectively. They are given by $\rho(\theta)=1/4\pi$ 
for the $[(2s_{1/2})^2]$ configuration and 
$\rho(\theta)=3/4\pi\cdot(5/4\cdot\cos^4\theta+3/20)$ 
for the $[(1d_{5/2})^2]$ configuration. As we see in the figures, the
angular density for the ground state is close to that for the pure 
$[(1d_{5/2})^2]$ configuration, while the angular density for the
second 0$^+$ state is close to that for the pure 
$[(2s_{1/2})^2]$ configuration, being consistent with the calculated 
values for $P_{dd}$ and $P_{ss}$ listed in Table II. 

\begin{figure}
\includegraphics[scale=1.3,clip]{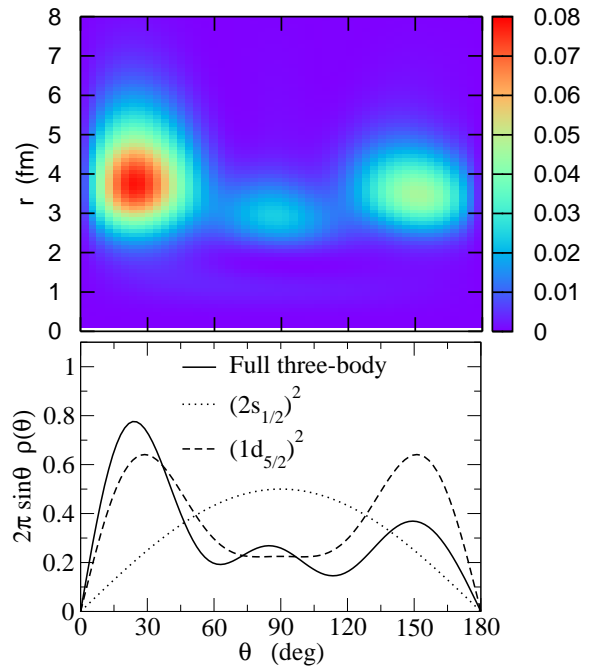}\\
\vspace{-1.45cm}
\hspace{-0.78cm}
\includegraphics[scale=0.415,clip]{fig1b}
\caption{(Color online) 
(the upper panel) The two-particle density for the ground state of $^{16}$C 
obtained with the parameter set D for the 
single-particle potential as a function of $r_1=r_2=r$ and 
the angle between the
valence neutrons, $\theta$.  It is weighted with a factor of 
$8\pi^2 r^4\sin\theta$. 
(the lower panel) The corresponding angular density weighted with a 
factor 2$\pi\sin\theta$. 
The solid line is the result of the three-body model calculation 
with the
Minnesota potential. The dotted and the dashed lines are for the pure 
$[(2s_{1/2})^2]$ and 
$[(1d_{5/2})^2]$ configurations, respectively. 
}
\end{figure}

\begin{figure}
\includegraphics[scale=1.3,clip]{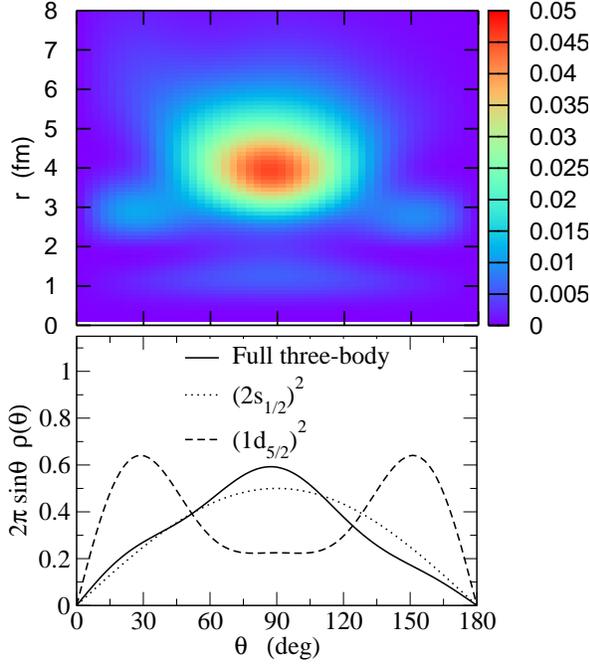}\\
\vspace {-1.45cm}
\hspace{-0.78cm}
\includegraphics[scale=0.415,clip]{fig2b}
\caption{(Color online) 
Same as Fig. 1, but for the second 0$^+$ state of $^{16}$C. 
}
\end{figure}

We next discuss the quadrupole excitation in $^{16}$C. Table III
summarizes the results of the present three-body model 
for the first 2$^+$ state.  The energy of the 2$^+$ state is well
reproduced with this model, especially with the parameter set D. 
As compared to the results of Refs. \cite{SMA04,HY06}, 
the probabilities for the $[2s_{1/2}1d_{5/2}]$ and $[(1d_{5/2})^2]$
components, denoted as $P_{sd}$ and $P_{dd}$, respectively, 
are comparable to each other in our calculation, whereas 
$P_{sd}$ is much larger than $P_{dd}$ in Ref. \cite{HY06}. 

\begin{table}[hbt]
\caption{Properties of the first 2$^+$ state of $^{16}$C obtained
with several parameter sets for the neutron-core potential, 
$V_{nC}$. $P_{sd}$ and $P_{dd}$ are the
probability for the $[(2s_{1/2})\times(1d_{5/2})]$ and 
$[(1d_{5/2})^2]$ components
in the wave function, respectively. $E_{2^+}$ is the excitation energy 
in the unit of MeV, while B(E2) is the electric quadrupole transition 
strength 
from the 2$^+$ state to the ground state, in the unit of $e^2$fm$^4$. 
The experimental values 
are $E_{2^+}$ (exp)=1.77 MeV and  B(E2; exp)=
0.63 $\pm$ 0.27 $e^2$fm$^4$ \cite{IOA04}, respectively. 
$e_{\rm pol}^{\rm I}$ is the core polarization charge which reproduces the 
experimental B(E2) value for the $^{15}$C nucleus within the
$n$-$^{14}$C model, whereas 
$e_{\rm pol}^{\rm II}$ takes into account the mass number dependence
according to Eq. (\ref{epol}). 
} 
\begin{center}
\begin{tabular}{c|ccc|cccc}
\hline
\hline
Set & $E_{2^+}$ & $P_{sd}$ & $P_{dd}$ & $e_{\rm pol}^{\rm I}$  
& B(E2; $e_{\rm pol}^{\rm I}$) & $e_{\rm pol}^{\rm II}$  
& B(E2; $e_{\rm pol}^{\rm II}$) \\
\hline
A & 1.26 & 0.392 & 0.504 & 0.162 & 0.972 & 0.145 & 0.808 \\
B & 1.33 & 0.400 & 0.515 & 0.160 & 0.937 & 0.144 & 0.781 \\
C & 1.34 & 0.402 & 0.500 & 0.153 & 0.956 & 0.137 & 0.797 \\
D & 1.63 & 0.472 & 0.406 & 0.122 & 1.074 & 0.109 & 0.899 \\
\hline
\hline
\end{tabular}
\end{center}
\end{table}

In order to calculate the E2 transition strength, we introduce the
core polarization charge, $e_{\rm pol}$. The E2 operator
$\hat{Q}_{2\mu}$ in the
present three-body model then reads (for $\mu$=0) \cite{SMA04,HY06}, 
\begin{eqnarray}
&&\hat{Q}_{20}=\left(\frac{Z_c}{A^2}\,e+\frac{A^2-2A+2}{A^2}\,e_{\rm
  pol}\right) \sum_{i=1}^2r_i^2Y_{20}(\hat{\vec{r}}_i) \nonumber \\
&&\hspace*{-0.7cm}
+\sqrt{\frac{5}{4\pi}}\left(\frac{Z_c}{A^2}\,e-2\frac{A-1}{A^2}\,e_{\rm
  pol}\right)
(2z_1z_2-x_1x_2-y_1y_2),
\end{eqnarray}
where $A=A_c+2$. The value of the core polarization charge which is
required to fit the experimental B(E2) value in the $^{15}$C nucleus, 
$B(E2; 5/2^+\to 1/2^+)$ = 0.97 $\pm$ 0.02 $e^2$ fm$^4$ \cite{AS91}, is
listed as $e_{\rm pol}^{\rm I}$ in the fifth column in Table III. 
Notice that 
these are significantly smaller
than that obtained with the harmonic vibration model of Bohr and
Mottelson, $e_{\rm BM}=0.55$ for $^{16}$C, which, however,  
does not include the effect of loosely-bound wave functions 
(see Eq. (6-386b) in Ref. \cite{BM75}). 
For a loosely-bound state, the polarization charge may be modified
 as 
\begin{equation} 
e_{\rm pol}=e_{\rm BM}\,\frac{3R^2/5}{\langle l_2j_2|r^2|l_1j_1\rangle}, 
\end{equation}
where $R=1.2 A^{1/3}$ fm and
 $\langle l_2j_2|r^2|l_1j_1\rangle$ is the radial matrix element
between the s.p. states ($l_1 j_1$) and ($l_2 j_2$)
(See Eq. (6-387) in Ref. \cite{BM75}). 
With the set D, we obtain the ratio $\frac{3R^2}{5}/
\langle 1d_{5/2}|r^2|2s_{1/2}\rangle$=0.205 as a reduction factor 
for the polarization charge 
for the transition from 1$d_{5/2}$ to 2$s_{1/2}$ s.p. states. This leads
to $e_{\rm pol}$=0.113, which is consistent with $e_{\rm pol}^{\rm I}$
shown in Table I. 
A similar small value of polarization charge 
has been obtained also with the self-consistent Hartree-Fock (HF) +
particle-vibration model \cite{SA01}. 
The calculated B(E2)
value for $^{16}$C with $e_{\rm pol}^{\rm I}$ 
is listed in the
sixth column in Table III. 
In contrast to the 
previous calculations with the three-body model \cite{SMA04,HY06},
which overestimated the B(E2) value for $^{16}$C with 
$e_{\rm pol}^{\rm I}$, our calculations reproduce well the experimental
B(E2) value. 
We notice that the small value of $P_{ss}$ and $P_{sd}$ in our 
wave functions is responsible for  good agreement with the experimental
 B(E2) value. For the parameter set D, 
the E2 matrix elements between  various two-particle configurations are
estimated to be, 
\begin{eqnarray}
&&\langle 2s_{1/2}\,1d_{5/2}|\hat{Q}_{20}|(2s_{1/2})^2\rangle =
-1.087~~e\,{\rm fm^2}, \\
&&\langle 2s_{1/2}\,1d_{5/2}|\hat{Q}_{20}|[(1d_{5/2})^2]^{J=0}\rangle =
-0.627~~e\,{\rm fm^2}, \\
&&\langle [(1d_{5/2})^2]^{J=2}|\hat{Q}_{20}|
[(1d_{5/2})^2]^{J=0}\rangle = -0.811 e\,{\rm fm^2}. 
\end{eqnarray}
Thus, the largest matrix element is 
 the one between the (2$s_{1/2})^2$ configuration in the 
ground state and the [2$s_{1/2}$\,1$d_{5/2}$] configuration in the
2$^+$ state, although the other two matrix elements 
have  substantial contributions. 
Naturally, a small value of $P_{ss}$ and $P_{sd}$ leads to a small 
B(E2) value, which is desired in order to reproduce the experimental
data. 
A further improvement of the calculated value of B(E2) can be achieved
if the mass number dependence of polarization charge is taken into
account. In Ref. \cite{SSH03}, the result of the HF+particle-vibration
model for the core polarization charge of carbon isotopes 
was parameterized as, 
\begin{equation}
e_{\rm pol}=0.82\frac{Z}{A}-0.25\frac{N-Z}{A}+\left(
0.12-0.36\frac{Z}{A}\,\frac{N-Z}{A}\right)\tau_z. 
\label{epol}
\end{equation}
This formula leads to the ratio of $e_{\rm pol}$ for $^{16}$C to
that for $^{15}$C to be $e_{\rm pol}(^{16}$C$)/e_{\rm pol}(^{15}$C)
=0.897. The polarization charge which is scaled by this
factor from $e_{\rm pol}^{\rm I}$ is denoted by $e_{\rm pol}^{\rm II}$
in Table III. We find that the calculated B(E2) values with 
$e_{\rm pol}^{\rm II}$ agree remarkably well with the experimental value within
the experimental uncertainty. 

We now discuss the longitudinal 
momentum distribution of $^{15}$C fragment in the breakup reaction of 
$^{16}$C nucleus. For this purpose, we calculate the 
the stripping cross section in the eikonal approximation
\cite{HM85,HBE96,E96,BH04}. That is \cite{SMA04,E96,BH04}, 
\begin{eqnarray}
&&\frac{d\sigma_{-n}}{dp_z}=\frac{1}{2\pi\hbar}\frac{1}{2l+1}
\sum_m\int^\infty_0d^2b_n\,[1-|S_n(b_n)|^2]
\nonumber \\
&&\hspace*{-0.6cm}
\times\int^\infty_0d^2r_\perp
|S_c(b_c)|^2 
\left|\int^\infty_{-\infty}dz\,e^{-ip_zz/\hbar}g_{lj}(r)Y_{lm}(\hat{\vec{r}})
\right|^2, 
\label{eikonal}
\end{eqnarray}
where $g_{lj}(r)$ is the radial part of 
the spectroscopic amplitude given by  
$\langle \Psi_{ljm}(^{15}{\rm C})|\Psi_{gs}(^{16}{\rm C})
\rangle=g_{jl}(r){\cal Y}_{jl-m}(\hat{\vec{r}})$ with 
${\cal Y}_{jl-m}(\hat{\vec{r}})$ being the spinor spherical
harmonics. 
$\vec{b}_n$ and $\vec{b}_c$ are the impact parameters for the neutron
and the core nucleus, respectively. They are related to the relative
coordinate between the neutron and the core nucleus,
$\vec{r}=(\vec{r}_\perp,z)$, by $\vec{b}_n=\vec{b}_c+\vec{r}_\perp$. 
In the eikonal approximation, the S-matrix is calculated as 
$S(b)=\exp(2i\chi(b))$ with 
\begin{equation}
\chi(b)=-\frac{1}{2\hbar v}\int^\infty_{-\infty}dz\,V(b,z),
\end{equation}
where $v$ is the incident velocity and $V(b,z)$ is an optical
potential between a fragment and the target nucleus. 

\begin{figure}
\includegraphics[scale=0.4,clip]{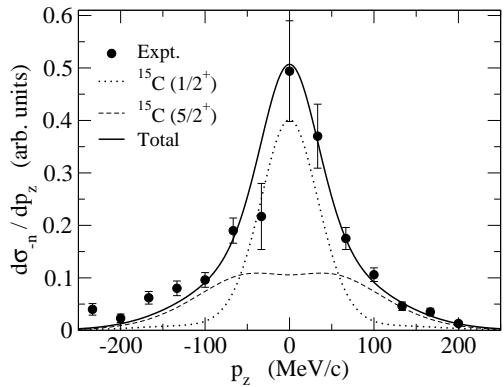}
\caption{The longitudinal momentum distribution of the 
$^{15}$C fragment from the breakup reaction of $^{16}$C on 
$^{12}$C target at 83 MeV/nucleon, obtained with the parameter set D. 
The contribution from the 1/2$^+$ and 5/2$^+$ states of $^{15}$C 
are shown by the dotted and the dashed lines, respectively. 
The experimental data are taken from Ref. \cite{Y03}. 
}
\end{figure}

Figure 3 compares 
the
eikonal approximation for the breakup reaction 
$^{16}$C + $^{12}$C $\to ^{15}$C + $X$ at $E=83$ MeV/nucleon 
with the experimental data \cite{Y03}.  
We use the optical potential of Comfort and Karp \cite{CK80} for the 
neutron-$^{12}$C potential. 
The optical potential between the $^{15}$C fragment and the target is
constructed with the single folding procedure using the $^{14}$C
density given in Ref. \cite{PZV00} and a s.p. wave function for the
valence neutron for a specified final state of the fragment nucleus. 
As is often done, we assume that the 
cross sections for diffractive breakup (i.e.,
elastic breakup) behave exactly the same as the
stripping cross sections as a function of longitudinal momentum, 
and thus we  scale the calculated cross section  (\ref{eikonal})
to match with the peak of the experimental data.
The dotted and dashed lines in Fig. 3 are contribution from 
the 1$s_{1/2}$ and 2$d_{5/2}$ states of the fragment $^{15}$C nucleus,
respectively. These are added incoherently to obtain the total 
one neutron removal cross section, which is denoted by the solid
line. Our result 
reproduces remarkably well the experimental 
longitudinal momentum distribution of the 
$^{15}$C fragment in the range of $-$200MeV/c $\le $p$_z \le$ 200MeV/c.

In summary, we have applied the $n$-$n$-$^{14}$C three-body model in
order to investigate the properties of the $^{16}$C nucleus. 
We diagonalized the three-body Hamiltonian with the finite range
Minnesota potential for the interaction between the valence neutrons. 
As the basis states,  we adopted the single-particle states obtained from
the Woods-Saxon potentials, in which the continuum spectrum is
discretized within the  large box. 
With this model, the experimental data for 
the root-mean-square radius, the B(E2) value from the
first 2$^+$ state to the ground state, and the longitudinal momentum
distribution of the $^{15}$C fragment from $^{16}$C breakup are all
reproduced well. In particular, we have succeeded to reproduce
the B(E2) value for both $^{15}$C and $^{16}$C nuclei 
simultaneously using the  core polarization charge which is consistent
with the one obtained  
with the particle-vibration coupling models. 
The calculated probability of the 
(1d$_{5/2})^2$ configuration in the ground state wave function of
$^{16}$C is about 70\% while that 
of the (2s$_{1/2})^2$ configuration is about 18\%. These values 
are close to those extracted from the 
analyses of the experimental longitudinal momentum distribution. 

\medskip

We thank W. Horiuchi, C.A. Bertulani, and N. Vinh Mau 
for useful discussions. 
We also thank T. Yamaguchi for sending us the experimental data in a
numerical form. 
This work was supported by the Japanese
Ministry of Education, Culture, Sports, Science and Technology
by Grant-in-Aid for Scientific Research under
the program numbers (C(2)) 16540259 and 16740139.

\end{document}